\documentclass[aip,amsmath,amssymb,reprint,twocolumn,nofootinbib]{revtex4-1} 
\pdfoutput=1
\usepackage[english]{babel}
\usepackage[utf8]{inputenc}
\usepackage{float}
\usepackage{graphicx}
\usepackage[abs]{overpic}
\usepackage{transparent}
\usepackage[caption=false]{subfig}
\usepackage{bm}
\usepackage{mathtools}
\usepackage{xcolor,varwidth}
\usepackage{siunitx}
\usepackage{dcolumn}
\usepackage{multirow}
\usepackage{booktabs}
\usepackage{array}
\usepackage[symbol]{footmisc}
\usepackage[]{natbib}

\setcitestyle{super}

\DeclareRobustCommand*{\citen}[1]{%
	\begingroup
	\romannumeral-`\x 
	\setcitestyle{numbers}%
	\cite{#1}%
	\endgroup   
}

\usepackage{silence}
\usepackage{color}
\newcommand{\MB}[1]{{\color{black} #1}} 
\newcommand{\NU}[1]{{\color{black} #1}} 
\newcommand{\TP}[1]{{\color{black} #1}} 
\newcommand{\MS}[1]{{\color{black} #1}} 

\makeatletter

\begin{document}

\title{Correction of beam hardening in X-ray radiograms} 

\author{\MB{Manuel Baur}}
\affiliation{Institute for Multiscale Simulation, Friedrich-Alexander-Universität, Nägelsbachstrasse 49b, 91052 Erlangen, Germany}
\author{\NU{Norman Uhlmann}}
\affiliation{Fraunhofer Institute for Integrated Circuits, Flugplatzstrasse 75, 90768 Fürth, Germany}
\author{\TP{Thorsten Pöschel}}
\affiliation{Institute for Multiscale Simulation, Friedrich-Alexander-Universität, Nägelsbachstrasse 49b, 91052 Erlangen, Germany}
\author{\MS{Matthias Schröter}}
\email{matthias.schroeter@ds.mpg.de}
\affiliation{Institute for Multiscale Simulation, Friedrich-Alexander-Universität, Nägelsbachstrasse 49b, 91052 Erlangen, Germany}
\affiliation{Max Planck Institute for Dynamics and Self-Organization, 37077 G\"ottingen, Germany}

\date{\today}


\begin{abstract}
The intensity of a {\it monochromatic} X-ray beam decreases exponentially with the distance it has traveled inside a material;
this behavior is commonly referred to as Beer-Lambert's law. Knowledge of the material-specific attenuation coefficient $\mu$
allows to determine the thickness of a sample from the intensity decrease the beam has experienced. 
However, classical X-ray tubes emit a {\it polychromatic} bremsstrahlung-spectrum. 
And the attenuation coefficients of all materials depend on the photon energy: photons with high energy are attenuated 
less than photons with low energy.  In consequence, the X-ray spectrum changes while traveling through the medium; due to the relative 
increase of high energy photons this effect is called beam hardening. For this varying spectrum, the Beer-Lambert law only remains valid 
if $\mu$ is replaced  by an \textit{effective} attenuation coefficient $\mu_\text{eff}$ which depends not only on the material, but also its 
thickness $x$ and the details of the X-ray setup used. We present here a way to deduce $\mu_\text{eff}(x)$ from a small number of auxiliary
measurements using a phenomenological model. This model can then be used to determine an unknown material thickness or   
in the case of a granular media its volume fraction.
\end{abstract}


\maketitle 


\section{Introduction}
If an X-ray photon travels through a material there exists for each atom it encounters a finite probability
that it will either be scattered inelastically at one of its electrons, 
or that it will be absorbed by kicking an electron out of the hull of the atom \cite{als-nielsen:11,pair_production}.
These probabilities themselves will depend on both the energy $E$ of the photon and
the type of atoms the material is made of, which is normally quantified by the atomic number $Z$.
In consequence, if a monoenergetic X-ray beam passes through a material, its intensity $I$ decreases exponentially 
with the distance $x$ traveled inside the sample:
\begin{equation}
I(x) = I_0 \exp(-\mu(E, Z) x).
\label{Equ:Beer-Lambert}
\end{equation}
\noindent
In this so-called Beer-Lambert's law $I_0$ is the intensity of the initial beam and $\mu$ is the attenuation coefficient which depends
on $E$ and $Z$. 

By measuring the ratio of intensities $I/I_0$ the thickness  $x$ of the material can be determined.
In granular systems this method can be used to determine the average volume fraction $\phi = x/L$ along the path of the photons where $L$ is the size of the container\cite{kabla:05,ribiere:07}. 
Because the temporal resolution  of this method is only limited by the frame rate of the detector, it can also be used to study dynamic systems such as granular flow \cite{michalowski:84,baxter_pattern_1989,kabla_dilatancy_2009}, 
impact \cite{royer_formation_2005,homan:15}, vertically vibrated samples \cite{cao:14},
liquid jets \cite{macphee:02}, the subsurface swimming of sandfisch lizards \cite{maladen:09},
fluidized beds \cite{kumar_gamma-ray_1995,shollenberger_gamma-densitometry_1997,mudde_gamma_1999,mudde_double_2010},
and two-phase flow \cite{makiharju_time-resolved_2013}.

\begin{figure}[b]
	\centering
	\includegraphics[width=\linewidth]{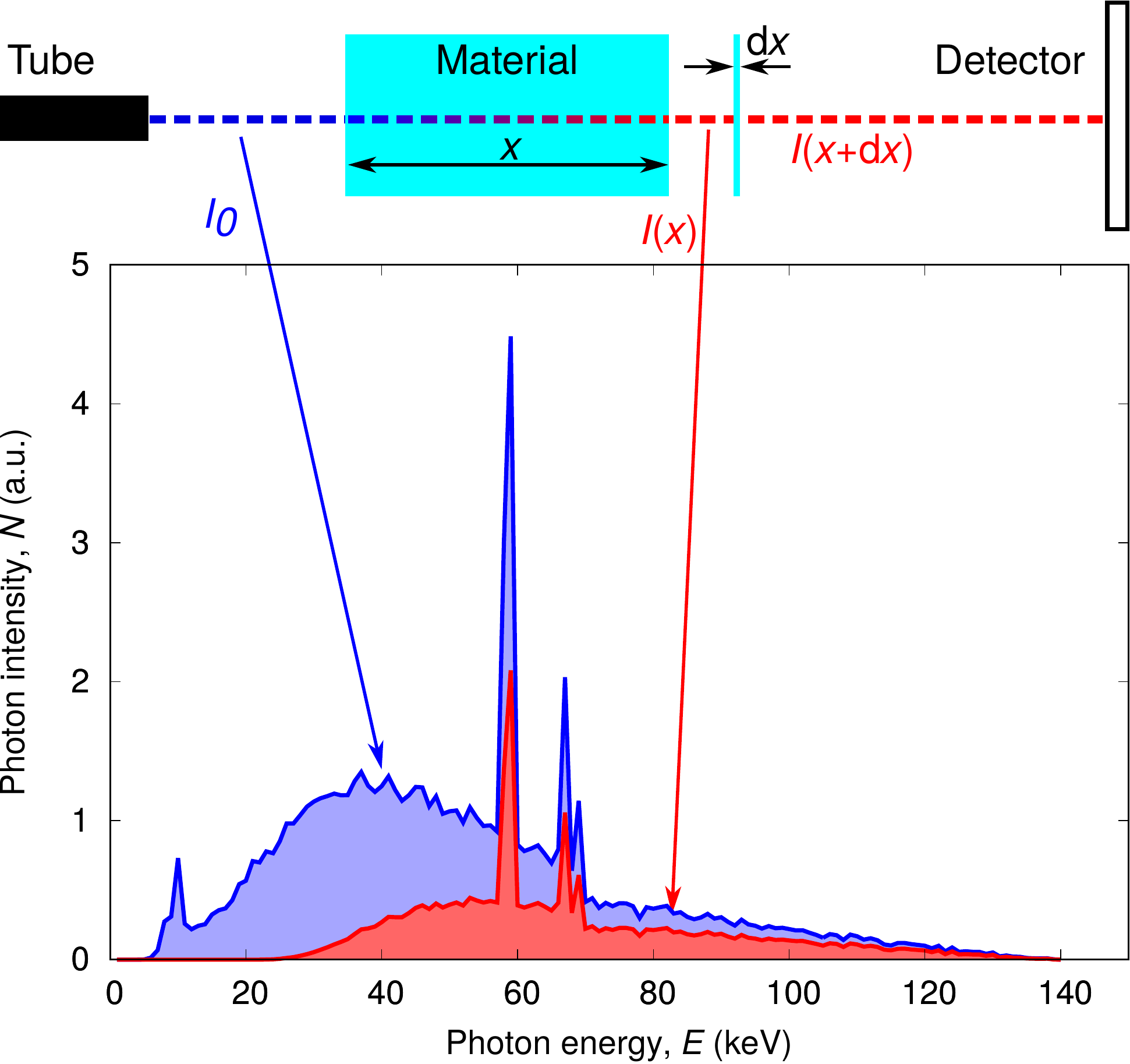}
	\caption{Beam hardening: The energy spectrum of an incident (blue) X-ray beam changes to the red spectrum while passing through an aluminum block of thickness $x$ =  1cm. Adding an infinitesimal thin slice of material $\mathrm{d}x$ does not change the red spectrum. The blue spectrum is created by a Monte Carlo simulation of a Comet MXR-225 X-ray tube (tungsten target, 140 kV acceleration voltage).}
	\label{Fig:Beam-hardening}
\end{figure}

 Classical X-ray tubes, which are normally used in scientific, industrial or medical setups, emit a broad energy spectrum originating 
mostly from the so-called bremsstrahlung.  Inside the material, low energy photons are attenuated stronger than high energy photons. In consequence, the relative contribution of the high energy part of the spectrum increases with material thickness, 
as shown in Fig.~\ref{Fig:Beam-hardening}. This process is known as beam hardening, its most immediate consequence
is that Eq.\ \ref{Equ:Beer-Lambert} is no longer applicable. In order to quantify the thickness of a material using its 
X-ray attenuation, beam hardening has to be taken into account.

One way to reduce the effect of beam hardening is the use of a filter, an additional sheet of material (typically a metal such as copper or aluminum) between the X-ray tube and the sample. This  
filter reduces the the ratio of low to high energy photons in the spectrum, i.e.~it is effectively narrowing the spectrum. However, this comes at the price of a reduced overall intensity of the beam; without completely removing the problem.

Another option is to replace the X-ray tube with a monoenergetic source. This could either be
a synchrotron beamline with a monochromator crystal in the beam path \cite{als-nielsen:11}   or 
a $\gamma$-ray sources such as  $^{137}$Cs \cite{kumar_gamma-ray_1995,mudde_gamma_1999,ribiere:07,shollenberger_gamma-densitometry_1997}. 
However, the first solution suffers from the small sample area of typically 1 cm$^2$ and the necessity to secure beam time at an user facility via a proposal. 
In the second approach, the high energies of the $\gamma$-rays result in a low contrast for many interesting samples
in fields such as soft matter and fluid dynamics. 

In many experimental situations the use of a polyenergetic photon spectrum cannot be avoided. In 
consequence, Eq.\ \ref{Equ:Beer-Lambert} has to be adapted by the use of an effective attenuation coefficient which depends on the material thickness. 

In this work we measure the effective attenuation coefficients for several materials on two different X-ray CT-setups. We describe all our data with a new heuristic model function which is shown to be more accurate than the models previously used in literature. We demonstrate how the thickness of a material can be deduced  from our model and quantify the occurring error.

\section{Attenuation of X-rays}\label{Sec:Background}
The intensity of an X-ray beam inside a material decreases due to two processes: 
scattering and absorption of the photons at the electrons of the material. 
Because the probability of a photon interacting with an electron depends 
only on the  energies of the photon and the electron, 
for monoenergetic photons this process is independent of the depth inside the medium. 
Therefore every slice of thickness $\mathrm{d}x$ attenuates the intensity $I$ by the same fraction: 
$\mathrm{d}I/I = - \mu \mathrm{d}x$, 
where the attenuation coefficient $\mu$ is a material parameter.
Integration leads to the Beer-Lambert law shown in Eq.\ \ref{Equ:Beer-Lambert}.

The attenuation coefficient $\mu(Z,\rho)$ depends on both the electron configuration of the atoms constituting 
the material (here summarized by the atomic number $Z$ of the elements) and on the density $\rho$ of the material.  The former dependence has been precomputed 
by the National Institute of Standards and Technology (NIST) and can be downloaded from their website using the online tool XCOM \cite{berger_nist_2010}.
The $\rho$ dependence is linear in the number of atoms per volume; 
considering this dependence explicitly with the so called mass attenuation coefficient $\mu/\rho$
simplifies the handling of mixtures, molecules, and compressible materials such as gases.

\begin{figure}[t]
	\centering
	\includegraphics[width=\linewidth]{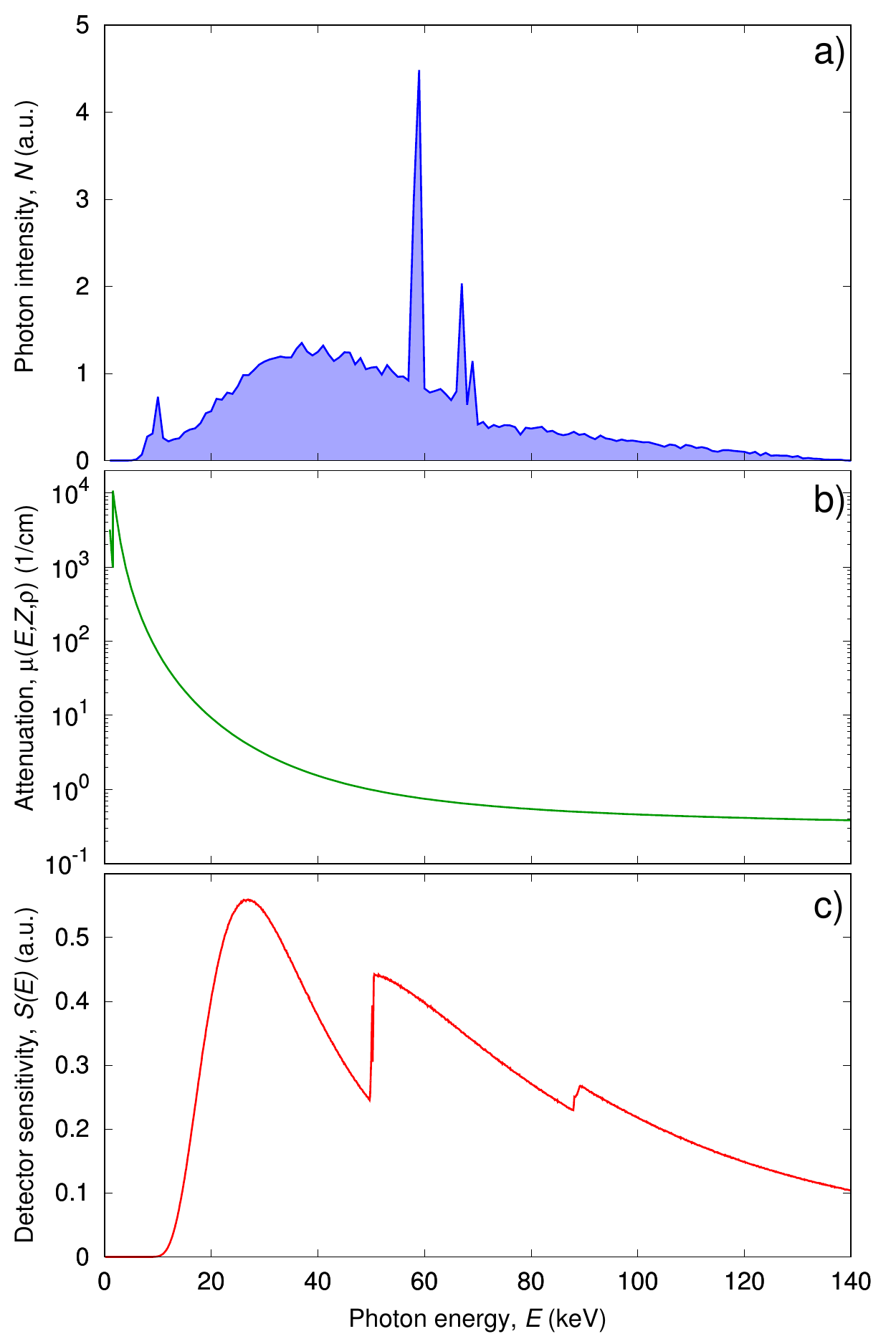}
	\caption{All three, the photon intensity of the source, the attenuation 
	coefficient of the sample, and the sensitivity of the X-ray detector, 
	depend on the energy of the
	X-ray photons. a) shows the results of a Monte Carlo simulation of a 
	Comet MXR-225 X-ray tube
    with a tungsten anode and an acceleration voltage of $140~\si{keV}$.
	b) is the attenuation coefficient of aluminum, retrieved from the XCOM database supplied by 
	 NIST \cite{berger_nist_2010}. 
	 c) describes the sensitivity of a Perkin Elmer XRD 820 AN 14 detector. This data was generated by simulating the detector 
	 using ROSI \cite{giersch:03,giersch:08}.
	 }
	\label{Fig:Int-Integral}
\end{figure}

For polychromatic beams photons of energy $E$ are attenuated according to their own attenuation coefficient  $\mu=\mu(E,Z,\rho)$. Due to the overall decrease of $\mu$ with $E$ (cf.~Fig.\ \ref{Fig:Int-Integral} b), 
the ratio of high to low energy photons increases
while the polychromatic beam passes through a material, as shown in Fig.\ \ref{Fig:Beam-hardening}. 
In consequence, the  attenuation of a polyenergetic X-ray spectrum is not 
described by the standard Beer-Lambert equation.

In normal X-ray imaging setups the beam intensity is measured by a detector, which 
responds to photons of different energies according to its spectral sensitivity  $S(E)$. 
Therefore the grayvalue of any given pixel will depend on three different factors:
the emitted X-ray spectrum $N(E)$, the attenuation coefficient $\mu(E,Z,\rho)$, 
the sensitivity curve  $S(E)$, and the material thickness $x$:
\begin{equation}
I(x) \propto \int N(E) \, \,  \exp\{-\mu(E,Z,\rho) \, x\}  \, \,  S(E) \, \, \mathrm{d}E
\label{Equ:Int-Integral}
\end{equation}

An example for the energy dependence of $N$, $\mu$, and $S$ is shown in Fig.\ \ref{Fig:Int-Integral}.
The problem with Eq.\ \ref{Equ:Int-Integral} is that most users will neither 
know $N(E)$ and $S(E)$ of their X-ray setup, nor will they have the means to measure these two curves. Therefore 
the tabulated values of $\mu(E)$ are insufficient to determine the material thickness from 
the measured intensity.

\subsection{Energy averaged attenuation coefficients}
For the description of polychromatic X-ray beams
two types of energy averaged attenuation coefficients are used in the literature:
a differential attenuation coefficient $\bar{\mu}$  and an integral versions  $\mu_\text{eff}$  
 \cite{bjarngard_attenuation_1994,yu_linear_1997,kleinschmidt_analytical_1999,alles_beam_2007,pease_monitoring_2012}.
While only  $\mu_\text{eff}$ can be measured in experiments, theoretical models have been developed for
both versions; we will therefore start by reviewing their relation.

The differential attenuation coefficient $\bar{\mu}(x)$ describes the intensity change
at a given depth $x$ inside the material \cite{kleinschmidt_analytical_1999}, 
it is defined by:

\begin{equation}
\frac{\mathrm{d} I(x)}{\mathrm{d} x} = -\bar{\mu}(x) I(x).
\label{Equ:Diff_mu_definition}
\end{equation}

Measuring  $\bar{\mu}(x)$ directly would require measuring the change 
in beam intensity $I(x+\mathrm{d}x) - I(x)$ due to an infinitesimally thin slice of material $\mathrm{d}x$, 
as indicated in Fig.\ \ref{Fig:Beam-hardening}. Which is in practice not feasible.  
One of the main applications of  $\bar{\mu}(x)$  is to calculate the absorbed energy dose in medical applications\cite{kleinschmidt_analytical_1999}.

In order to describe how the beam intensity decreases inside the medium from $I_0$ to $I(x)$, we have to integrate 
Eq.\ \ref{Equ:Diff_mu_definition}:
\begin{equation}
I(x) = I_0 \exp\left(-\int_{0}^{x} \bar{\mu}(x') \mathrm{d}x'\right).
\label{Equ:Diff-Beer-Lambert}
\end{equation}

The differential attenuation coefficient, $\bar{\mu}(x)$ can however be measured indirectly:
Eq.\ \ref{Equ:Diff-Beer-Lambert} can be rewritten as
\begin{equation}
-\ln \frac{I(x)}{I_0} = \int_{0}^{x} \bar{\mu}(x') \mathrm{d}x'.
\end{equation}

\begin{figure}[t]
	\includegraphics[width=\linewidth]{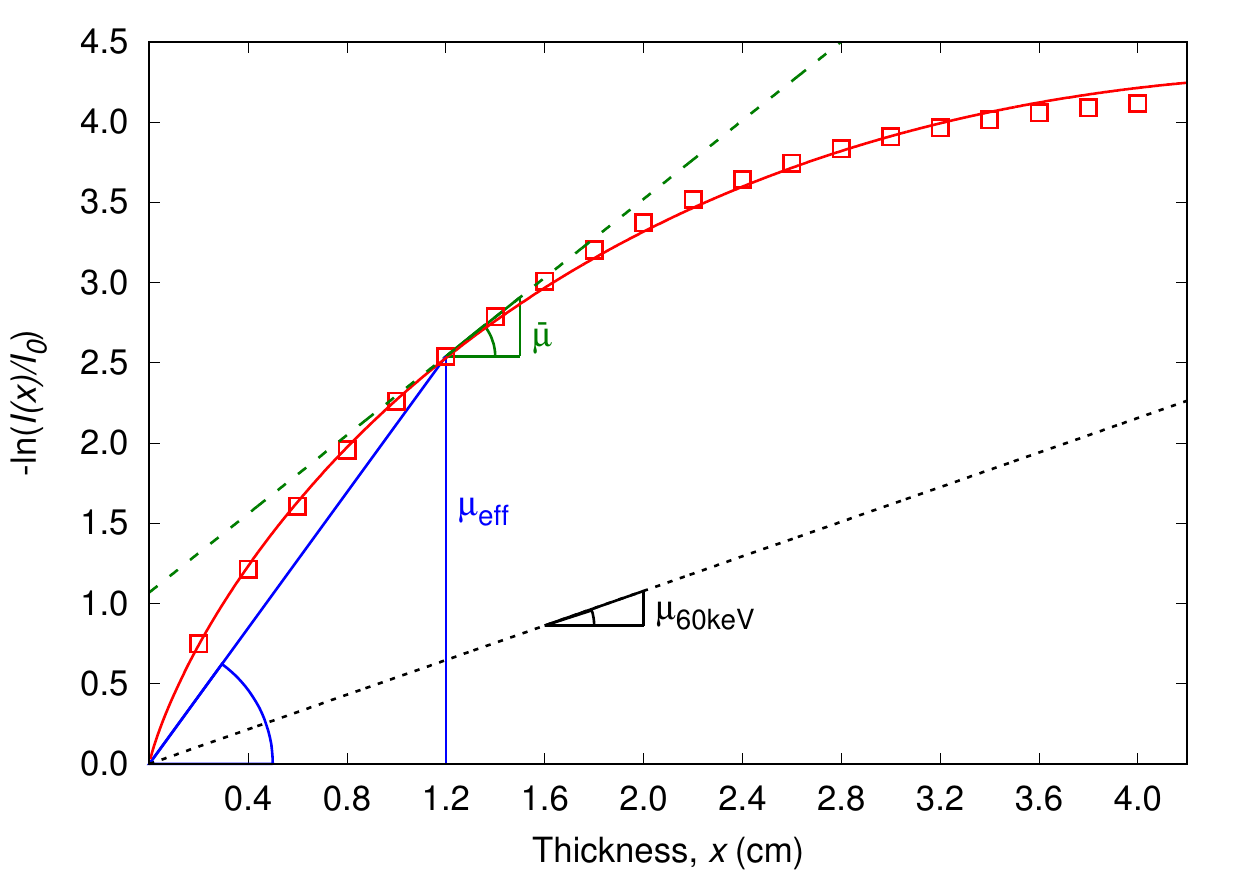}
	\caption{A graphical comparison of $\bar{\mu}$ and $\mu_\text{eff}$, both for an $x$ value of 1.2 cm
	(following ref.~\citen{pease_monitoring_2012}).
	The experimental data (red squares) correspond to the logarithm of the
    intensity decrease of an X-ray beam ($60~\si{kV}$, tungsten anode) passing through borosilicate glass
    slabs of  different thickness.
    A smooth representation of the data is obtained by a fit with Eq.\ \ref{Equ:OurFit} (red line). 
    $\bar{\mu}$ in the depth 1.2 cm is the slope of the tangent to the fit curve at $x=1.2~\si{cm}$ 
    (green dashed line, Eq.\ \ref{eq:mu_bar_image}).
	 $\mu_\text{eff}$ of a glass slab of thickness 1.2 cm is the slope of the blue secant 
	 connecting the origin with the data point at $x=1.2~\si{cm}$  (Eq.\ \ref{Equ:Comp_mus_mu_eff}). 
	 Finally, $\mu_\text{60keV}$ indicates the attenuation a monochromatic beam of  
	 $60~\si{keV}$ energy would experience in borosilicate glass \cite{hubbell_tables_2009} (black short dashed line). 
	 Due to the monotonic decrease of $\mu(E)$, no amount of 
    beam hardening can result in a slope of $\mu_\text{eff}$ smaller than $\mu_\text{60keV}$.
	}
	\label{Fig:Comp_mus}
\end{figure}

By differentiation we obtain:
\begin{equation}
\frac{\mathrm{d}}{\mathrm{d}x} \left(- \ln \frac{I(x)}{I_0}\right) = \bar{\mu}(x).
\label{eq:mu_bar_image}
\end{equation}
This means that $\bar{\mu}(x)$ can be obtained as the slope of the tangent
if the data is plotted as in Fig.\ \ref{Fig:Comp_mus}.

In contrast to $\bar{\mu}(x)$, which describes the attenuation only at a certain depth inside the material,
the integral or effective attenuation coefficient $\mu_\text{eff}$  averages over the whole sample of thickness $x$ 
\cite{bjarngard_attenuation_1994, yu_linear_1997}:
\begin{equation}
I(x) = I_0 \exp(-\mu_\text{eff}(x) x).
\label{Equ:Eff-Beer-Lambert}
\end{equation}

In consequence, it is easy to determine single $\mu_\text{eff}$ values
by comparing the X-ray intensities before and after a material of known thickness. 
And if the functional dependence of $\mu_\text{eff}$ on the material thickness $x$ is known,
the width of an unknown object can be computed from a single radiogram, as shown in section \ref{Sec:Length_meas}.

Eq.\ \ref{Equ:Eff-Beer-Lambert} can be rewritten as
\begin{equation}
-\ln \frac{I(x)}{I_0} = \mu_\text{eff}(x) x.
\label{Equ:Comp_mus_mu_eff}
\end{equation}
This implies that in Fig.\ \ref{Fig:Comp_mus} $\mu_\text{eff}(x)$ can be visualized as the slope of the secant which connects the origin with the datapoint at thickness $x$. 

There exists a generic lower bound for $\mu_\text{eff}(x)$ provided  that the monoenergetic attenuation
coefficient $\mu$ is a monotonic decreasing function of the photon energy; which is normally the case in the 
experimentally relevant energy range.
A simple gedankenexperiment shows then that the effect of beam hardening will stop when the only photons remaining from the 
initial spectrum are the ones with the highest energy  $\mu(E_\text{max})$, i.e.~the acceleration voltage of 
the X-ray tube. Therefore $\mu_\text{eff}(x)$ has to be larger than $\mu(E_\text{max})$.

The conversion between the two types of attenuation coefficients is straightforward \cite{kleinschmidt_analytical_1999,pease_monitoring_2012}: 
By inserting Eq.~\ref{Equ:Eff-Beer-Lambert} into Eq.~\ref{Equ:Diff_mu_definition} we obtain:
\begin{equation}
\bar{\mu}(x) = \mu_\text{eff}(x) + \frac{\mathrm{d}\mu_\text{eff}(x)}{\mathrm{d}x} x.
\label{Equ:Transform_mus}
\end{equation}

The reverse relationship follows from Eq.\ \ref{Equ:Diff-Beer-Lambert} and \ref{Equ:Eff-Beer-Lambert}:
\begin{equation}
\mu_\text{eff}(x) = \frac{\int_0^x \bar{\mu}(x') \mathrm{d}x'}{x}.
\label{Equ:Transform_mus2}
\end{equation}


\subsection{Modeling of $\bm{\mu_\text{eff}}$}\label{Sec:Modeling}
Eq.\ \ref{Equ:Int-Integral} summarizes the essence of the effect of beam hardening: 
The measured intensity does not only depend on the type and thickness of the sample material, but also
on the shape of the initial beam spectrum $N(E)$ and the sensitivity curve of the detector $S(E)$.  
Most users of an X-ray setup have no information about the exact shape of $N(E)$ and $S(E)$ and 
in consequence the integral in Eq.\ \ref{Equ:Int-Integral} cannot be solved.

Given this situation, a number of models, either completely heuristic, or based on some physical arguments of $N(E)$,
have been suggested for $\bar{\mu}(x)$ and $\mu_\text{eff}(x)$. 
Because all of these models omit at least part of the physics contained in  Eq.\ \ref{Equ:Int-Integral}, 
their merit can only be assessed by comparing them with experimental data. We will present such a comparison, 
with focus on energy scales  and materials used in typical Computed Tomography X-ray setups,
in section \ref{Sec:Results}.

The first models for beam hardening were introduced by
Bjärngard \& Shackford \cite{bjarngard_attenuation_1994}, and Yu \textit{et al.} \cite{yu_linear_1997} 
in order to improve dose calculations for medical applications.
They gathered data  for water and aluminum at linear accelerators for radiotherapy 
at acceleration voltages of 6 MV and 25 MV and fitted $\mu_\text{eff}$ with:
\begin{align*}
	\mu_\text{eff}(x) &= \mu_0 - \lambda x 
	\tag{BS} \label{Equ:BS}\\
	\mu_\text{eff}(x) &= \frac{\mu_0}{1+\lambda x} 
	\tag{Yu 1} \label{Equ:Yu1}\\
	\mu_\text{eff}(x) &= \frac{\mu_0}{(1+\lambda x)^\beta} 
	\tag{Yu 2} \label{Equ:Yu2}
\end{align*}
where $\mu_0$, $\lambda$, and $\beta$ are all fit parameters.

Kleinschmidt \cite{kleinschmidt_analytical_1999,mix-up} computed $\bar{\mu}$ values from numerical data and suggested as
fit function:
\begin{equation}
\bar{\mu}=\mu(E_\text{max})+\frac{\mu_1}{1+\lambda_1 x+\lambda_2 x^2},
\label{Equ:Initial_KS}
\end{equation}
with $\mu_1$, $\lambda_1$ and $\lambda_2$ as free parameters. In order to compare 
 Eq.\ \ref{Equ:Initial_KS} with our experimental data, we
transform it using Eq.\ \ref{Equ:Transform_mus2} to:
\begin{multline*}
\mu_\text{eff}(x) = \mu(E_\text{max}) +
\frac{2 \mu_1}{x \sqrt{-\lambda_1^2 + 4 \lambda_2}} \times \qquad\qquad\quad \\
\left[
\arctan\left(
\frac{\lambda_1 + 2 \lambda_2 x}{\sqrt{-\lambda_1^2 + 4 \lambda_2}}
\right)
-\arctan\left(
\frac{\lambda_1}{\sqrt{-\lambda_1^2 + 4 \lambda_2}}
\right)
\right]
\tag{KS}
\label{Equ:KS}
\end{multline*}

Alles \& Mudde \cite{alles_beam_2007} derived an expression for $\bar{\mu}$, which contains ten summands 
in a compound fraction; it is based on four free parameters which have to be determined by fits to experimental data.
In later publications  Mudde and coworkers \cite{mudde_feasibility_2008,brouwer_effects_2012,gomez-hernandez_fast_2016} 
analyzed their data with a simpler expression for the intensity decay at the detector:
\begin{equation}
I/I_0 = A + B \exp(-x/C),
\label{Equ:Mu_intensity}
\end{equation}
where $A$, $B$ and $C$ are fit parameters.
For the comparison with our experimental data we combined Eq.\ \ref{Equ:Eff-Beer-Lambert} with Eq.~\ref{Equ:Mu_intensity}.
to obtain an equivalent $\mu_\text{eff} (x)$ as:
\begin{align*}
	\mu_\text{eff}(x) = -\frac{1}{x} \ln \left[A+B \exp(-x/C)\right] 
	\tag{MU} \label{Equ:MU}
\end{align*}

Another model, which is based on the Lambert-W function, was suggested by Mathieu \textit{et al.} \cite{mathieu_empirical_2011}.
However, its underlying assumption is not compatible with our experimental data, as shown in the appendix \ref{Sec:App_Model_Mathieu}. We will therefore not include it in our discussion.

Finally, we suggest here a purely heuristic model for $\mu_\text{eff}(x)$:
\begin{equation}
\mu_\text{eff}(x) = a + \frac{b}{x^\alpha},
\label{Equ:OurFit}
\end{equation}
where $a$, $b$ and $\alpha$ are free parameters. 
In section \ref{Sec:Results_Validation} we compare Eq.\ \ref{Equ:OurFit} to the former models, namely
Eq.s \ref{Equ:BS}, \ref{Equ:Yu1}, \ref{Equ:Yu2}, \ref{Equ:KS}, and \ref{Equ:MU}.

\section{Experimental procedure}\label{Sec:Exp_Procedure}
\begin{figure}[b]
	\centering
	\includegraphics[width=\linewidth]{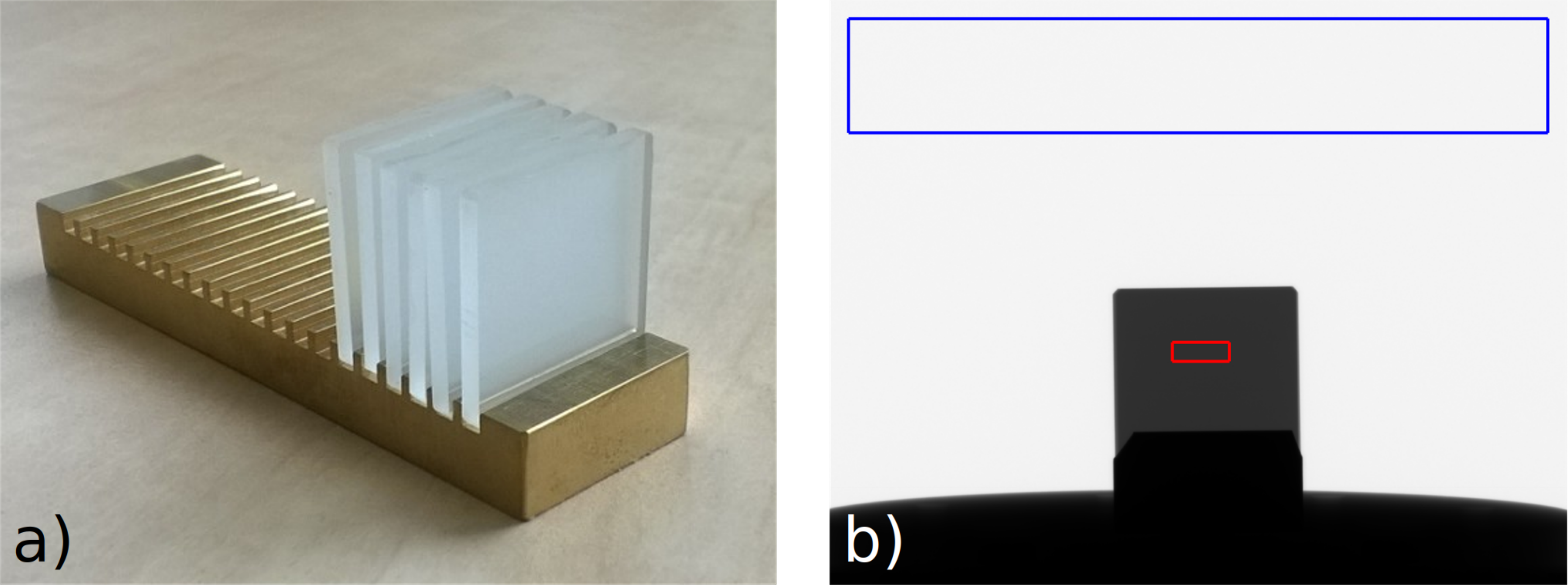}
	\caption{Creating borosilicate samples of varying material thickness. 
	a) glass plates of 2 mm thickness are stacked in a brass sample holder.
	b) radiogram of the sample. The intensities $I_0$ and $I(x)$ are measured in the  blue and red framed areas.
	}
	\label{Fig:plates_radiogram}
\end{figure}

In order to create samples with a well defined thickness $x$ in the range $2-40~\si{mm}$, we stack
up to 20 borosilicate plates of thickness 2 $\pm$0.05 mm in the self-made sample-holder shown in Fig.\ \ref{Fig:plates_radiogram} a). Samples are placed between an X-ray tube and a camera
and radiograms of the type shown in \ref{Fig:plates_radiogram} b) are captured. 
The effective attenuation coefficient $\mu_\text{eff}$ is then measured using eq.~\ref{Equ:Comp_mus_mu_eff}.
The values of $I(x)$ and $I_0$ are extracted from the radiograms; they correspond to the 
 mean gray values of regions where the beam is transmitted through respectively passes above the plates, 
as shown in Fig.\ \ref{Fig:plates_radiogram} b).

Measurements are performed in a standard X-ray tomograph typical for scientific and industrial applications.
It contains an X-ray-worx tube (XWT-160-TCHE Plus) with a tungsten transmission target and a PerkinElmer DEXELA 1512 14 bit flat panel detector \cite{Konstantinidis_2012}.
In section \ref{Sec:Results_2nd_Setup} we compare the $\mu_\text{eff}$ values computed with this setup 
with values gathered from a second setup using a different camera and source.

The intensity of an X-ray beam is mainly diminished by two effects: photoelectric absorption and Compton scattering. 
Some fraction of the scattered photons will also hit the detector, just not at the position
predicted by geometrical optics. This contribution to the image intensity will not only depend on the X-ray spectrum
and the sample material, but also on the sample shape; in general it will be impractical to predict it. 
However, if the geometry of the test samples used to quantify beam hardening  resembles the geometry of the 
samples used in the actual measurements, the effect of scattering will be captured by the heuristic approximation 
presented here. 

Absorption measurements are often performed using wedges \cite{bjarngard_attenuation_1994} and step wedges \cite{kinds_feasibility_2011} as single samples providing a variety of material thicknesses. However, because
wedges have a broken spatial symmetry in direction perpendicular to the X-ray beam, 
the undefined scattering contributions limit the accuracy of such measurements. 

In contrast, the plate stack configuration shown in Fig.\ \ref{Fig:plates_radiogram} a) is symmetric with respect to the center beam. Together with a small geometric magnification this reduces the contribution of scattering to an 
area in the vicinity of the rim of the plates. The area in the center of the plates, where $I$ is measured, can be chosen 
such that it is free of spatial gradients \cite{yu_linear_1997}. The obvious disadvantage of the plate stack is the requirement
of a larger number of individual measurements. However, we will show in section \ref{Sec:Results_Error} that three measurements are sufficient.


\section{Comparison model and experiment}\label{Sec:Results}
In this section we present measurements  of the effective attenuation coefficient $\mu_\text{eff}$, as defined in Eq.\ \ref{Equ:Eff-Beer-Lambert}, for different acceleration voltages of the X-ray tube and sample materials. We also 
describe how $\mu_\text{eff}$  changes due to pre-filtering of the beam and using another camera and X-ray tube. 
All experimental data can be described with Eq.\ \ref{Equ:OurFit} which is also shown to be more accurate than the 
other models introduced in section \ref{Sec:Modeling}.

\begin{figure}[t]
	\centering
	\includegraphics[width=\linewidth]
		{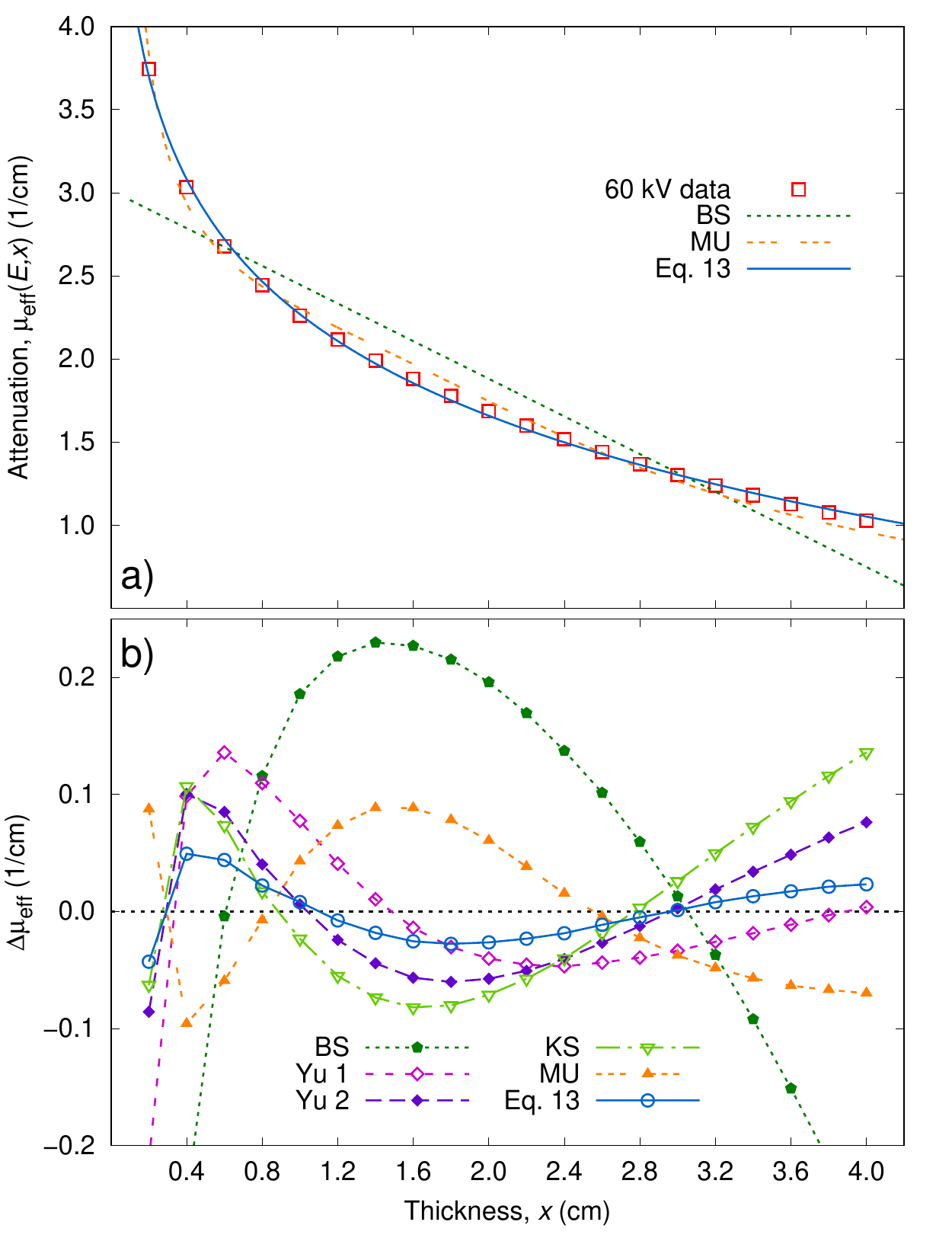}
	\caption{Low energy comparison of the different models for $\mu_\text{eff}$, demonstrating that 
    Eq.\ \ref{Equ:OurFit} provides the best fit to the experimental data.
	a) For visual clarity we fit only a subset of the model functions 
	to experimental data gathered for borosilicate glass plates 
    at an acceleration voltage of 60 kV. 
	b) $\Delta \mu_\text{eff}$ is the difference between a given model and the experimental data. 
	The figure includes all fit-functions discussed  in section \ref{Sec:Modeling}; for  the  model acronyms see there.
	Lines in panel a) are fits to the  measured data, in panel b) lines are guides to the eye.
	 }
	\label{Fig:mu_eff_comp_KS_60kV}
\end{figure}

\begin{figure}[t]
	\centering
	\includegraphics[width=\linewidth]
		{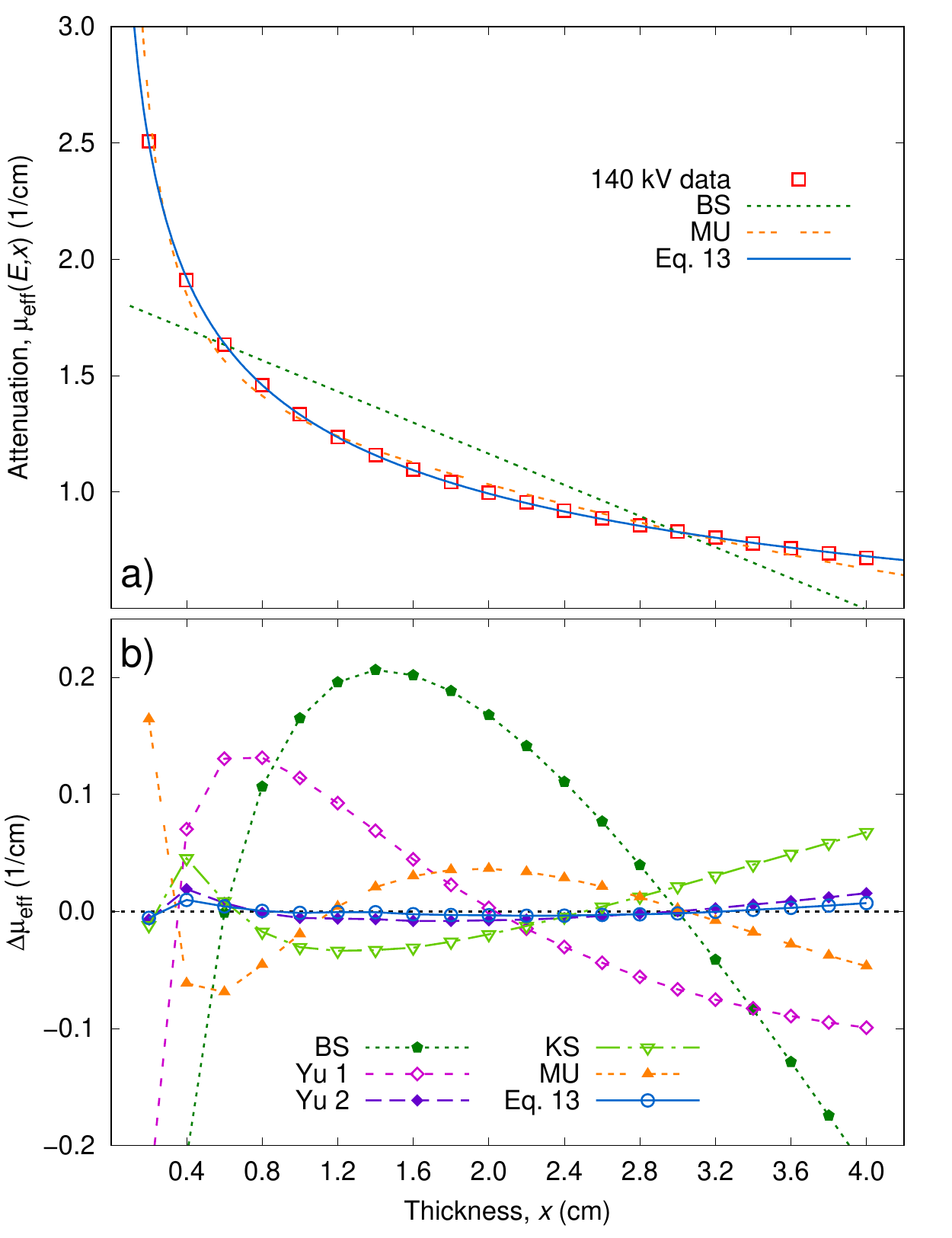}
	\caption{High energy comparison of the different models for $\mu_\text{eff}$. 
	The sample is borosilicate glass plates, the acceleration voltage 140 kV. 
	The composition of the figure is analog to Fig.\ \ref{Fig:mu_eff_comp_KS_60kV}; 
	model acronyms are defined in section \ref{Sec:Modeling}.
	Eq.\ \ref{Equ:OurFit} provides again the best fit.
	}
	\label{Fig:mu_eff_comp_KS_140kV}
\end{figure}

\subsection{Comparison of the different models}\label{Sec:Results_Validation}
When optimizing an X-ray imaging setup, one needs to choose the optimal acceleration voltage:
Lower energies deliver typically a stronger contrast between different materials, as it is e.g.~beneficial for 
composite or soft materials. Higher energies result in a lower effective attenuation and therefore the possibility
to image thicker or denser samples. 

In order to cover both cases, we present in Figs.\ \ref{Fig:mu_eff_comp_KS_60kV} and \ref{Fig:mu_eff_comp_KS_140kV} measurements for $\mu_\text{eff}$ performed with 60 kV and 140 kV acceleration voltage.  
Both figures compare the experimental results with fits of the models discussed in section \ref{Sec:Modeling}.
For reasons of readability  the direct comparisons (Fig.\ \ref{Fig:mu_eff_comp_KS_60kV}a and 
\ref{Fig:mu_eff_comp_KS_140kV}a) do not include all models; but the plots of the $\mu_\text{eff}$ differences 
between model and data (Figs.\ \ref{Fig:mu_eff_comp_KS_60kV}b and \ref{Fig:mu_eff_comp_KS_140kV}b) do include them all.

The main result of Figs.\ \ref{Fig:mu_eff_comp_KS_60kV} and \ref{Fig:mu_eff_comp_KS_140kV} is that
Eq.\ \ref{Equ:OurFit} provides the best fit at both acceleration voltages. 
At 60 kV our model deviates less than $0.05~\si{1/cm}$ from the experimental $\mu_\text{eff}$
over the full range of sample thickness studied, at 140 kV less than $0.01~\si{1/cm}$.

The rather poor performance of model Eq.\ \ref{Equ:BS} is not surprising given that it was developed for much higher X-ray energy of 6 MV and 25 MV transmitted through water, which is a weakly attenuating medium. 
The model Eq.\ \ref{Equ:Yu2} is the second best choice; similar to our model it has an exponent as a free parameter. 
Model Eq.\ \ref{Equ:MU} did not converge to a reasonable approximation when fitted to  $\mu_\text{eff}$
\cite{info_gnuplot}. We therefore fitted Eq.\ \ref{Equ:Mu_intensity} directly to the $I(x)/I_0$  values and converted the result to $\mu_\text{eff}$ via Eq.\ \ref{Equ:Eff-Beer-Lambert}. 

\subsection{Effect of pre-filtering}
\begin{figure}[t]
	\includegraphics[width=\linewidth]{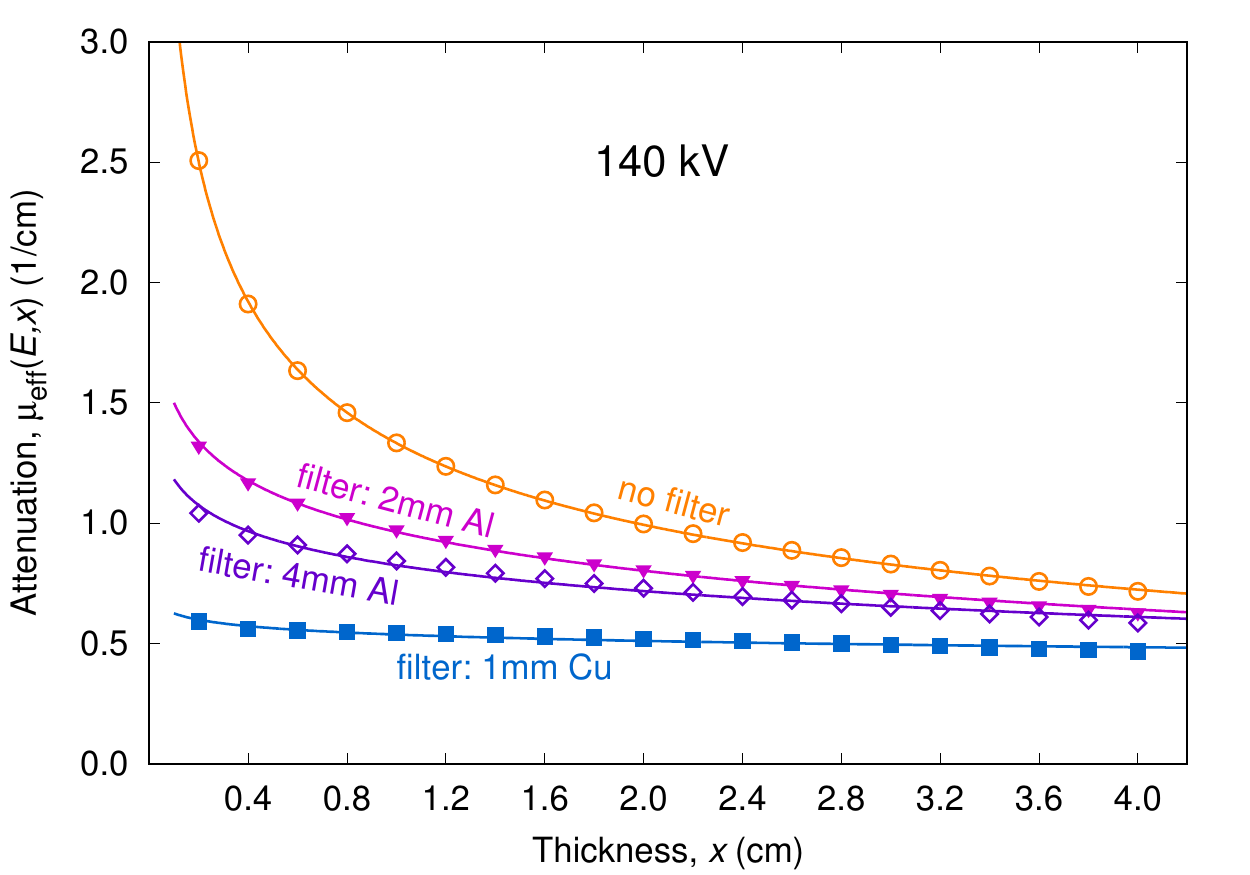}
	\caption{Adding a filter in the beam path decreases beam hardening.
	Experimental data are gathered for borosilicate glass and  140 kV acceleration voltage.
    Going from no filter, via  $2~\si{mm}$ and $4~\si{mm}$ aluminum filter, to a $1~\si{mm}$ copper filter 
    not only decreases the values of $\mu_\text{eff}$, it also decreased their dependence on $x$, 
    i.e.~the beam hardening effect.
    Eq.\ \ref{Equ:OurFit} provides a good fit for all applied filters.
    } 
	\label{Fig:mu_eff_bglass_filters}
\end{figure}

As described in the introduction, a common method to reduce the effect of beam hardening is to insert a small metal
plate into the beam path, directly in front of the X-ray tube. These metal filters remove more photons
from the low energy part of the spectrum, effectively narrowing the range of energies in the beam.
This leads indeed to a decrease in beam hardening as shown in Fig.\ \ref{Fig:mu_eff_bglass_filters}:
The dependence  of  $\mu_\text{eff}$ on $x$ decreases with increasing thickness and increasing atomic number 
of the filter inserted into the beam. A second effect is that the values of $\mu_\text{eff}$ also decrease
the more the spectrum is shifted towards the high energy range, in agreement with Fig.\ \ref{Fig:Int-Integral} b.
Most important in our context is however that Eq.\ \ref{Equ:OurFit} continues to provide a good model
for the experimental data, independent of the applied filter.

The major disadvantage of filtering is invisible in Fig.\ \ref{Fig:mu_eff_bglass_filters}: the narrowing in
the energy spectrum is accompanied by an overall decrease in intensity. For our measurements 
$I_0$ decreases to 65\% of the unfiltered intensity when adding the 2mm Al filter. For the 4mm Al filter this number
becomes 48\%, and for the  1mm Cu only 23 \% of the unfiltered intensity remains.

\subsection{Material independence of Eq.\ \ref{Equ:OurFit}}

\begin{figure}[t]
	\includegraphics[width=\linewidth]{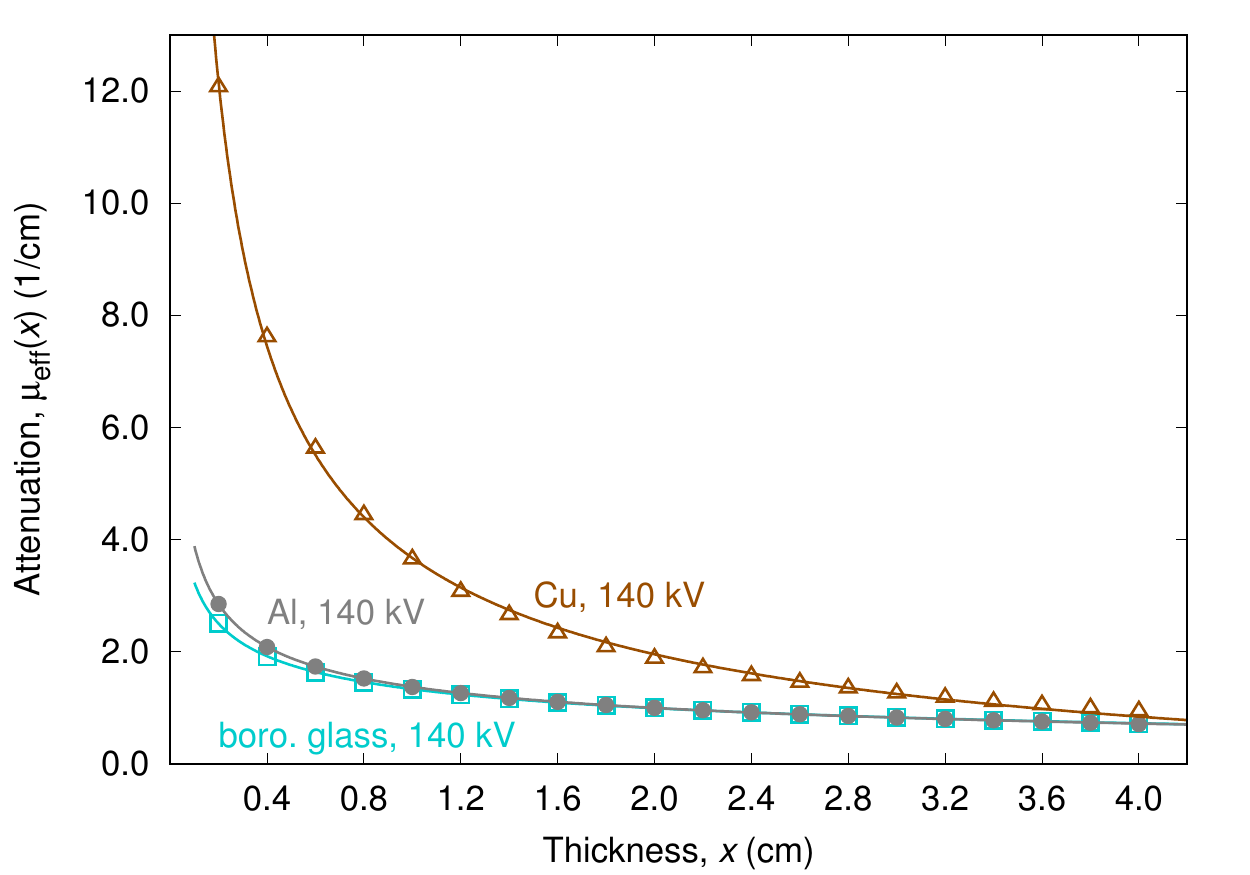}
	\caption{Eq.\ \ref{Equ:OurFit}  (solid lines) provides a good fit for	
	$\mu_\text{eff} (x)$ measurements of borosilicate glass (squares), aluminum (circles) and copper (triangles).
	All experimental data are measured at an acceleration voltage of 140 kV.
	}
	\label{Fig:mu_eff_materials}
\end{figure}

Fig.\ \ref{Fig:mu_eff_materials} demonstrates that our model (Eq.\ \ref{Equ:OurFit}) 
provides a good fit to $\mu_\text{eff} (x)$ for a variety of different materials. As in the case for
the borosilicate glass, the measurements 
for copper and aluminum are made  with stacks of 2 mm thick plates, cf.~Fig.\ \ref{Fig:plates_radiogram}.

The absolute values of $\mu_\text{eff} (x)$ require some explanation. 
For small values of $x$, copper attenuates much stronger than borosilicate glass and aluminum,
which is in  agreement with the higher atomic number of copper. However, for larger values of $x$
all three materials approach similar values of $\mu_\text{eff}$. This seems to imply that aluminum and copper
would be equally appropriate choices for shielding against X-rays; which is objectively not the case.
This apparent similarity of $\mu_\text{eff}$ is an artifact resulting from the limitations of our experimental setup:
the intensity $I$ behind the material decreases so much that the dark field noise of the camera starts to 
become a significant part of the signal, and this dark field noise is obviously not material-dependent.
A second spurious contribution to $I$ comes from photons scattered in the air and other parts of the setup.
This underlines again that $\mu_\text{eff}(x)$ is not a material property alone, but also 
dependent on the details of the experimental setup. However, within these limits fits with
Eq.\ \ref{Equ:OurFit} describe the data well and provide therefore the opportunity to compute 
the material thickness $x$ from measured values of $I$ and $I_0$; which is our actual goal.

\subsection{Device independence of Eq.\ \ref{Equ:OurFit}}\label{Sec:Results_2nd_Setup}
\begin{figure}[t]
\includegraphics[width=\linewidth]{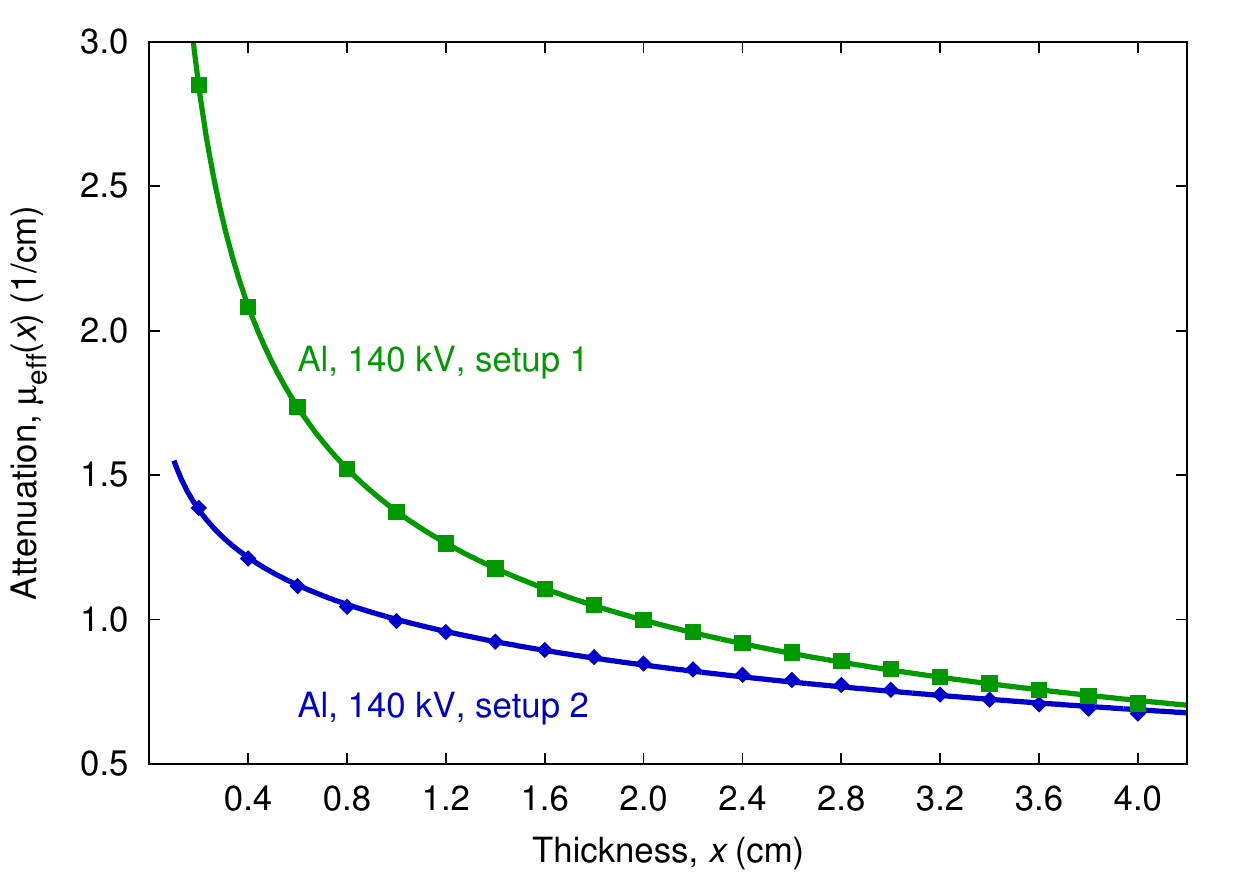}
\caption{While $\mu_\text{eff}(x)$ does depend on the details of the experimental setup, 
Eq. \ref{Equ:OurFit} continues to provide a good description (solid lines) of the data (squares and diamonds). 
Setup 1 consists of the  X-ray-worx tube and PerkinElmer detector combination used in the remainder of the paper, 
setup 2 combines a GE HighPower X-ray tube with an XEye detector. Both measurements are performed with an acceleration voltage of 140 keV.
}
\label{Fig:mu_eff_2nd_setup}
\end{figure}

As shown in Eq.\ \ref{Equ:Int-Integral}, the measured intensity $I(x)$ and therefore also  
$\mu_\text{eff}(x)$ will depend on the type of X-ray tube and detector used in a given setup. 
This is demonstrated in fig.\ \ref{Fig:mu_eff_2nd_setup} which compares two data sets: the  aluminum data already
shown in Fig.\ \ref{Fig:mu_eff_materials} and another data set captured 
with the a setup consisting of a 
GE 225HP 225kV HighPower X-ray tube and an XEye 2020 detector with a 300 $\mu$m thick CsI scintillator.
Even for the same acceleration voltage, the absolute values of $\mu_\text{eff}(x)$
differ up to the factor of two. However, both data set are again well described 
by a fit with Eq.\ \ref{Equ:OurFit}. 

\section{Determining the material thickness $\bm{x}$}\label{Sec:Length_meas}
Aim of this work is to measure the thickness $x$ of a material based on the intensity values extracted from a radiogram. 
As shown in section \ref{Sec:Results}, equation \ref{Equ:OurFit} provides the best known approximation 
of the effective absorption coefficient $\mu_\text{eff} (x)$.

Because $\mu_\text{eff} (x)$ depends on details of the setup such as acceleration voltage and camera type,
we need to calibrate our setup/material combination in addition to the actual measurement.
I.e.~we need to determine the three parameters $a$, $b$ and $\alpha$ in equation \ref{Equ:OurFit} 
by taking radiograms of objects of known thickness $x$ and made from the material we are interested in,
using the same setup we then use for the actual measurements. 
We will show in subsection \ref{Sec:Results_Error} that three calibration measurements, which can e.g.~be gathered from a scalene cuboid made of the sample material, are sufficient to determine  $a$, $b$ and $\alpha$.

The actual measurement of $x$ consists of determining the gray values $I$ and $I_0$ at  positions in the radiogram where the beam has either traveled through or passed next to the object (cf. Fig.\ \ref{Fig:plates_radiogram}).
Rewriting Eq.\ \ref{Equ:Eff-Beer-Lambert} we obtain
\begin{equation}
- \ln \left(\frac{I} {I_0}\right) = \mu_\text{eff}(x) x
\label{Equ:measure}
\end{equation}
inserting $\mu_\text{eff}(x)$ from Eq.\ \ref{Equ:OurFit} leads to
\begin{equation}
a x + b x^{1-\alpha} + \ln\left(\frac{I}{I_0}\right) = 0.
\label{Equ:Root}
\end{equation}

Because Eq.\ \ref{Equ:Root} cannot be solved analytically for $x$, we have to determine the material thickness indirectly:
We can either compute a look-up table for the right hand side of equation \ref{Equ:measure} and interpolate $x$ with the desired accuracy. Or we can solve Eq.\ \ref{Equ:Root} numerically, using e.g.~Newton's method. The latter method will
converge, if $x$ is restricted to the range $0 < x \leq x_\text{max}$ where $x_\text{max}$
is the maximal thickness of the sample. $x=0$ needs to be excluded, because Eq.\ \ref{Equ:OurFit} diverges at that point.

\subsection{Number of calibration measurements required}\label{Sec:Results_Error}

\begin{figure}[t]
	\includegraphics[width=\linewidth]{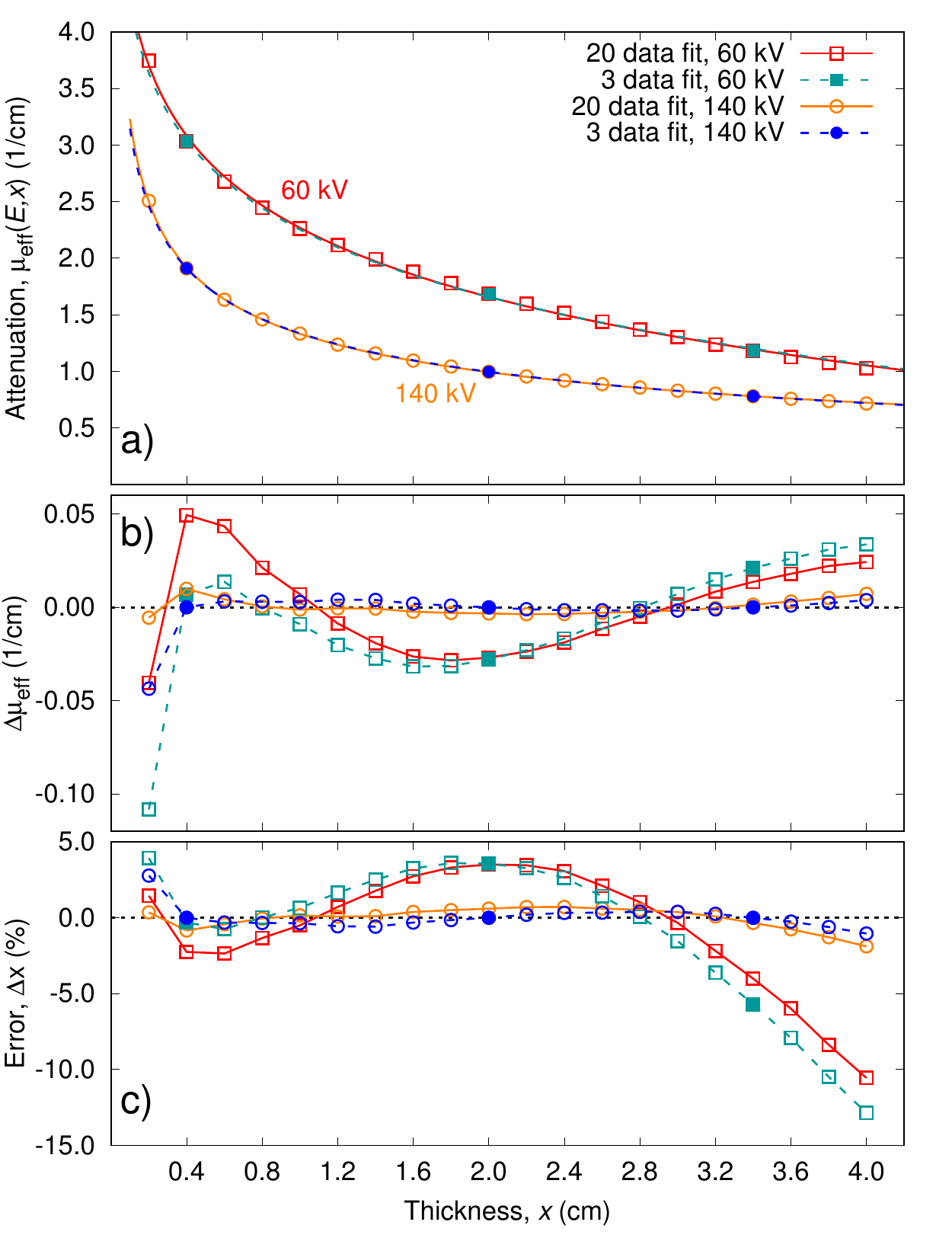}
	\caption{
	The number of calibration measurements required to determine $\mu_\text{eff}$ can be reduced to three with just a small loss of accuracy.
	a) squares and circles represent our experimentally determined $\mu_\text{eff}$ values  of borosilicate glass plates 
	for acceleration voltages of 60 kV and 140 kV. The solid lines are fits of Eq.\ \ref{Equ:OurFit} to all 20 data points (open symbols), the dashed lines to 3 data points (closed symbols). The two fits are hard to distinguish. Therefore we
	plot in panel b) $\Delta \mu_\text{eff}$ which is the difference between the experimental data and the fits to 20, respectively 3 data points. 
	c) Based on the two   $\mu_\text{eff} (x)$ fit curves, we can use the experimentally measured intensities to predict 
	the sample thickness $x$. From our knowledge of the actual sample thickness we can then determine the relative error
	$\Delta x$. (The lines in panel a) represent equation \ref{Equ:OurFit}, in panel b) and c) they are only 
    guides to the eye.)}
	\label{Fig:Length_meas_3datapoints}
\end{figure}

Section \ref{Sec:Results} demonstrates that Eq.\ \ref{Equ:OurFit} provides a good fit to our data, 
provided the fit is based on 20 data points. 
However, performing 20 calibrations measurements with different sample thicknesses is often not practicable. 
Figure \ref{Fig:Length_meas_3datapoints} shows that 3 calibration measurements suffice to 
determine $x$ with almost identical accuracy, as expected for an equation with three free parameters.
The red squares and orange circles in panel \ref{Fig:Length_meas_3datapoints} a) 
show the two borosilicate glass measurement series for 60 kV and 140 kV
acceleration voltage which were already discussed in section \ref{Sec:Results_Validation}. 
The data for each acceleration voltage is fitted twice with Eq.\ \ref{Equ:OurFit}: the solid lines are
fits to all 20 data points, the dashed lines are fits to only the three data points marked with filled symbols. 
In both cases the two fits are right on top of each other.

In panel \ref{Fig:Length_meas_3datapoints} b) we compare the 
absolute differences between the measured $\mu_\text{eff}$ values and the fits. 
At 140 kV the difference between the 3 and the 20 point fits is very small; even a small
extrapolation beyond the minimum and maximum thickness of the three sample points is possible.    
At 60 kV the difference between the two fits becomes more pronounced, 
but it is still of the same order as the deviation between the fits and the actually measured $\mu_\text{eff}$ values.

The actual aim of our measurements is the determination of the unknown sample thickness $x$.  
Fig.\ \ref{Fig:Length_meas_3datapoints} c)
shows the relative error $\Delta x$ of a measurement using the two fits, where we have solved 
Eq.\ \ref{Equ:Root} using  Newton's method. 
$\Delta x$ is below 15\% for the 60 kV and below 3\% for the 140 kV measurement; the differences between the 3 point and the 20 point fit are again small.

\section{Example: measuring the volume fraction in a granular shear band}

\begin{figure*}[h]
	\includegraphics[width=\linewidth]{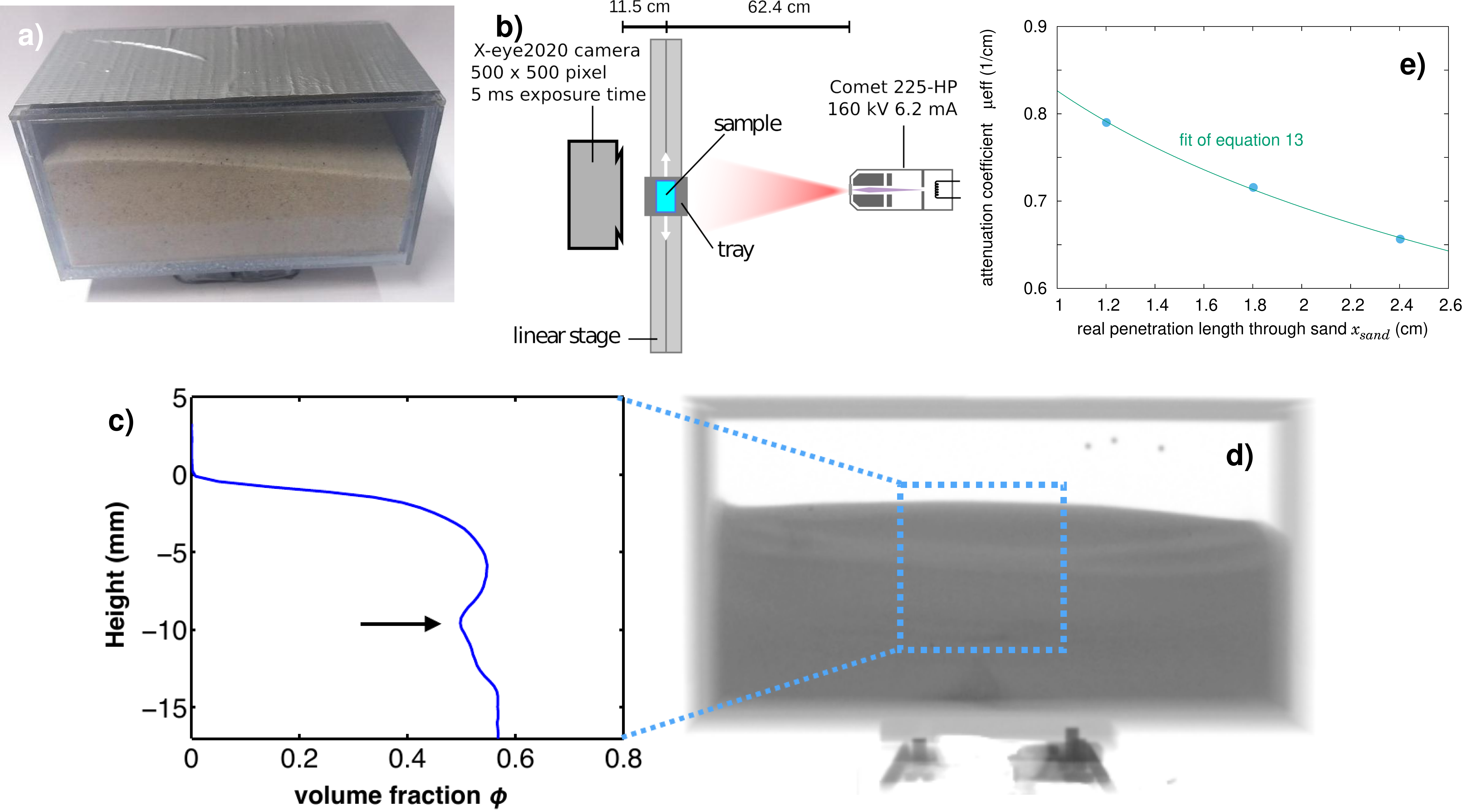}
	\caption{Measuring the volume fraction in granular shear bands.
	a) A polycarbonate box with inner dimensions 100 by 50 by
      50 mm$^3$  is filled with sand with a mean diameter of
      265$\pm$70 $\mu$m.
    b) Radiograms are taken while the box is shaken horizontally
    on a linear translation stage 
    with a frequency of 18 Hz.
    From the radiograms (panel d) the average volume fraction inside the sample 
    can be computed (panel c), provided $\mu_\text{eff} (x)$ of the sand is known (panel f).
    The latter was determined from attenuation measurements of a box with 
    side lengths 2, 3, and 4 cm which was filled with sand at a volume fraction 
    of 0.6.
    For further  information see \cite{kollmer:19}.
	}
	\label{Fig:example}
\end{figure*}

When dense granular systems are sheared, the strain is often localized
in so-called shear bands. One way to create such a shear band is to fill a 
rectangular box with sand while maintaining a free surface, cf. figure \ref{Fig:example} a).
If the box is then shaken horizontally, the upper part of the sample material sloshes back and forth between 
the outer walls while the bottom part of the sample moves stationary with the box; 
between these two parts a shear band forms.

Using high-speed X-ray radiography we can show that the formation of these shear bands is 
accompanied by dilatancy, i.e.~a reduction in the volume fraction $\phi$ (which measures the
locally averaged ratio of particle volume to total volume).  
Figure \ref{Fig:example} b) shows the corresponding setup: an X-ray beam is traveling perpendicular 
to the shaking direction through the sample cell. The corresponding radiogram (figure \ref{Fig:example} d)
displays brighter horizontal stripes; these correspond to the shear bands with their lower value of $\phi$.   

For a quantitative analysis of the radiograms shown in figure \ref{Fig:example} d) we need to first convert the
intensities in the radiogram to lengths $x_{\rm sand}$ that the X-rays travel through the actual sample material while passing through the box (here we assume that we can neglect the  $\mu_\text{eff}$ of the interstitial air). 
This step requires the knowledge of  $\mu_\text{eff} (x)$ of the sample material, using the method described in this
paper. Figure \ref{Fig:example} e) shows a fit of equation \ref{Equ:OurFit} to the intensity ratios measured
with an cuboidal box filled with sand of a known volume fraction.
The volume fraction averaged along the beam path can then be computed as
$\phi = x_{\rm sand} / L$, where $L$ is the inner wall to wall distance in beam direction.

Figure \ref{Fig:example} c) shows the volume fraction as a function of
height, measured and horizontally averaged inside the blue box in panel d), the arrow indicates the position 
of the shear band. Further information on the dynamics of these shear bands can be found in reference \cite{kollmer:19}.

\section{Conclusion}
In all X-ray imaging setups working with a broad energy spectrum, which is all setups using a classical 
X-ray tube, the attenuation has to be described by an effective attenuation coefficient
which does depend on both the type of material and its thickness. The latter dependence
originates from beam-hardening, the change of the energy spectrum within the material. 
Because both the intensity of the X-ray tube and the sensitivity of the detector 
are energy dependent, the properties of the experimental setup will 
influence how the effective attenuation depends on the sample thickness.
The new phenomenological equation for the effective attenuation 
introduced in this work provides a good fit to experimental
data gathered for a variety of materials and experimental conditions. 
It also allows reliable measurements of the sample thickness 
using as little as three calibration measurements to determine the effective attenuation. 
However, due to the large number of possible experimental configurations, a general applicability cannot be guaranteed.

\begin{acknowledgments}
We wish to thank Jonathan Kollmer for his help with Figure \ref{Fig:example}. This work was funded by the German Federal Ministry for Economic Affairs and Energy, grant number 50WM 1653.
\end{acknowledgments}

\appendix*

\section{Excluding the model by Mathieu \textit{et al.}}\label{Sec:App_Model_Mathieu}
\begin{figure}[t]
	\includegraphics[width=\linewidth]{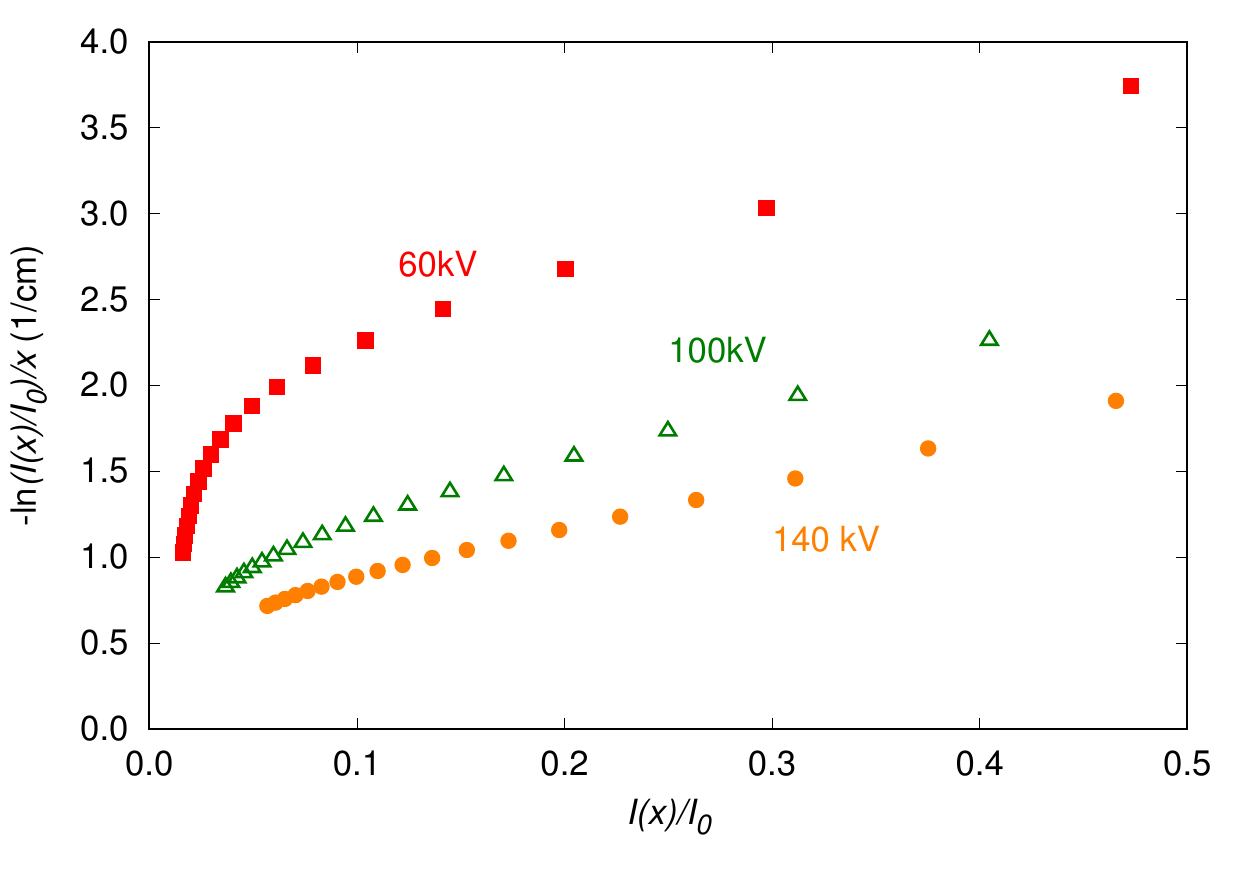}
	\caption{The model by Mathieu {\it et al.} does not describe our experimental data: Equation \ref{Equ:Mathieu}
	predicts the data of our borosilicate glass measurements to be on straight lines.
	}
	\label{Fig:LambertW}
\end{figure}

Mathieu and coworkers \cite{mathieu_empirical_2011} proposed an attenuation model based on the Lambert W function. 
According to this  model the measured intensities should be described by the following equation:
\begin{equation}
\frac{-\ln(I(x)/I_0)}{x} = \mu_0 + \lambda \frac{I(x)}{I_0},
\label{Equ:Mathieu}
\end{equation} 
However, our data clearly deviate from eq.~\ref{Equ:Mathieu} as shown in Fig.\ \ref{Fig:LambertW} where it would correspond to straight lines with slope $\lambda$.

\bibliography{beam_hardening_new}

\begin{thebibliography}{35}
\bibitem [{pai()}]{pair_production}%
  \BibitemOpen
  \href@noop {} {}\bibinfo {note} {If the energy of the x-ray photon exceeds
  1.022 MeV it can also be transformed into an electon-positron
  pair.}\BibitemShut {Stop}%
\end{thebibliography}

\end{document}